\documentclass[12pt,preprint]{aastex}
\usepackage{xspace}
\usepackage{graphicx}
\usepackage{natbib}
\usepackage{amssymb}
\usepackage{longtable}
\usepackage[caption=false]{subfig}

\newcommand{\hide}[1]{}

\newcommand{\gsim}{\ensuremath{\,\gtrsim\,}\xspace}
\newcommand{\lsim}{\ensuremath{\,\lesssim\,}\xspace}
\newcommand{\gl}{\ensuremath{\ell}\xspace}
\newcommand{\gb}{\ensuremath{{\it b}}\xspace}

\newcommand{\lb}{\ensuremath{(\gl,\gb)}\xspace}

\newcommand{\kms}{\ensuremath{\,{\rm km\,s^{-1}}}\xspace}

\newcommand{\K}{\ensuremath{\,{\rm K}}\xspace}

\newcommand{\percc}{\ensuremath{\,{\rm cm^{-3}}}\xspace}

\newcommand{\degree}{\ensuremath{\,^\circ}\xspace}

\newcommand{\msun}{\ensuremath{\,M_\odot}\xspace}     
 
\newcommand{\hii}{{\rm H\,{\footnotesize II}}\xspace}

\newcommand{\co} {\ensuremath{^{\rm 12}{\rm CO}}\xspace}
\newcommand{\cor}{\ensuremath{^{\rm 13}{\rm CO}}\xspace}

\slugcomment{lda, \today}
\shorttitle{CO Observations of RCW\,120}
\shortauthors{Anderson et al.}
\begin{document}

\title{Mopra CO Observations of the Bubble H{\small II} Region RCW\,120}

\author{L.~D.~Anderson\altaffilmark{1,2},
  Deharveng,~L.\altaffilmark{3}, Zavagno,~A.\altaffilmark{3},
  Tremblin,~P.\altaffilmark{4,5}, Lowe,~V.\altaffilmark{6,7},
  Cunningham,~M.~R.\altaffilmark{6}, Jones,~P.\altaffilmark{6},
  A.M.~Mullins\altaffilmark{8}, M.P.~Redman\altaffilmark{9}}

\altaffiltext{1}{Department of Physics and Astronomy, West Virginia University, Morgantown, WV, USA}
\altaffiltext{2}{Also an Adjunct Astronomer at the National Radio Astronomy Observatory, PO Box 2, Green
  Bank, WV 24944, USA}
\altaffiltext{3}{Aix Marseille Universit\'e, CNRS, LAM (Laboratoire d'Astrophysique de Marseille) UMR 7326, 
13388, Marseille, France}
\altaffiltext{4}{Laboratoire AIM Paris-Saclay (CEA/Irfu - Uni. Paris
  Diderot - CNRS/INSU), Centre d'etudes de Saclay, 91191
  Gif-Sur-Yvette, France}
\altaffiltext{5}{Astrophysics Group, University of Exeter, EX4 4QL Exeter, UK}
\altaffiltext{6}{School of Physics, University of New South Wales, NSW 2052, Australia}
\altaffiltext{7}{Australia Telescope National Facility, CSIRO Astronomy and Space Science, PO Box 76, Epping, NSW 1710, Australia}
\altaffiltext{8}{Irish Research Council, \emph{EMBARK} Scholar, at NUI Galway}
\altaffiltext{9}{Director, Centre for Astronomy, NUI Galway}

\begin{abstract}
We use the Mopra radio telescope to test for expansion of the
molecular gas associated with the bubble \hii\ region RCW\,120.  A
ring, or bubble, morphology is common for Galactic \hii\ regions, but
the three-dimensional geometry of such objects is still unclear.
Detected near- and far-side expansion of the associated molecular
material would be consistent with a three-dimensional spherical
object.  We map the J = $1\rightarrow 0$ transitions of \co,
  \cor, C$^{18}$O, and C$^{17}$O, and detect emission from all
  isotopologues.  We do not detect the $0_0 \rightarrow 1_{-1}E$
  masing lines of CH$_3$OH at 108.8939\,GHz.  The strongest CO emission is from the
photo-dissociation region (PDR), and there is a deficit of emission
toward the bubble interior.  We find no evidence for expansion of the
molecular material associated with RCW\,120 and therefore can make no
claims about its geometry.  The lack of detected expansion is roughly
in agreement with models for the time-evolution of an \hii\ region
like RCW\,120, and is consistent with an expansion speed of
  $\lsim 1.5\,\kms$.  Single-position CO spectra show signatures of
expansion, which underscores the importance of mapped spectra for such
work.  Dust temperature enhancements outside the PDR of RCW\,120
coincide with a deficit of emission in CO, confirming that these
temperature enhancements are due to holes in the RCW\,120 PDR.
H-alpha emission shows that RCW\,120 is leaking $\sim 5\%$ of the
ionizing photons into the interstellar medium (ISM) through PDR
holes at the locations of the temperature enhancements.  H$\alpha$
emission also shows a diffuse ``halo'' from leaked photons not
associated with discrete holes in the PDR.  Overall $\sim~25\pm10\%$
of all ionizing photons are leaking into the nearby ISM.

\end{abstract}


\section{Introduction}
{\it Spitzer}-GLIMPSE observations of the Galactic plane at
8.0\,\micron\ \citep{benjamin03} revealed the presence of $\sim\!600$
Galactic infrared (IR) rings, or ``bubbles'' seen in projection
\citep{churchwell06, churchwell07}.  \citet{simpson12} expanded this
to nearly 6000 objects, and using Widefield Infrared Survey Explorer
\citep[WISE;][]{wright10} data \citet{anderson14} cataloged more than
8000 objects that have this same basic morphology.  Nearly all bubbles
enclose \hii\ regions \citep{deharveng10}, and about half of all
\hii\ regions are bubbles \citep{anderson11}.  The common, circularly
symmetric bubble morphology for \hii\ regions indicates that
\hii\ regions expand almost uniformly.  The small morphological
differences between bubbles are likely caused by evolutionary effects
coupled with density variations in the ambient medium.

The 8.0\,\micron\ {\it Spitzer} emission traces the \hii\ region
photo-dissociation region (PDR) and surrounds the ionized gas.  It is
commonly assumed that the 8.0\,\micron\ emission is in part due
to polycyclic aromatic hydrocarbons (PAHs).  These molecules are
destroyed within the harsh radiation field of the \hii\ region itself
\citep{povich07, pavlyuchenkov13}, but fluoresce when hit with
ultra-violet photons in a weaker radiation field.  The
8.0\,\micron\ emission therefore traces \hii\ region PDRs.  It is this
emission at 8.0\,\micron\ from the \hii\ region PDR that forms the
bubble structure.  The PDR surrounds the ionized gas of the
\hii\ region, but it is frequently patchy, which allows some radiation
to escape.

Despite their ubiquity, the three-dimensional morphology of bubbles is
still unclear.  For 43 bubbles, \citet{beaumont10} found no evidence
for expansion, using \co\,($3\!\rightarrow\!2$) observations.  For a
three-dimensional geometry, associated molecular gas from the near-
and far-side PDRs of an expanding \hii\ region should be blue- and red-shifted,
respectively.  Their observations placed a limit on the bubble shell
thickness of generally $\lsim20\%$ of the bubble radius (their
Figure~2).  Furthermore, for some bubbles in their sample 
  even a thin three-dimensional shell does not agree with their data.
Because of these non-detections, \citet{beaumont10} concluded that
bubbles are two-dimensional rings formed in flat ($\gsim 10\,$pc
thick) parental molecular clouds.

Using {\it Herschel} observations, however, \citet{anderson12b} found
that 20\% of the total far-infrared emission of bubble \hii\ regions
comes from the direction of bubble ``interiors,'' the locations inside
the PDRs, which suggests a three-dimensional morphology. They found
that the percentage of far-infrared (FIR) emission coming from the bubble interiors
decreases with wavelength, from $26\pm0.7\%$ at 100\,\micron\ to
$12\pm7\%$ at 500\,\micron.  \citet{anderson11} found that the radio
recombination line width is no greater for bubbles than for the
\hii\ region population at large.  If bubbles were two-dimensional
structures, we would expect their line widths to be greater because
expansion out of the plane would not be restricted by dense
molecular material (unless all \hii\ regions are two-dimensional
  objects).  That broader lines are not detected for bubbles
argues for three-dimensional structures and/or low expansion velocities.

The geometry of bubbles is important for our understanding of
triggered star formation.  Numerous candidate triggered star formation
regions have been identified along bubble borders, suggesting that
triggering may be an important star formation process
\citep[e.g.,][]{thompson12, simpson12}.  Classical models of
\hii\ regions predict that, due to the pressure difference between the
warm ($\sim\,10,000$\,K) ionized gas and the surrounding cold neutral
material ($\sim\,10\,$K), \hii\ regions expand with characteristic
velocities of a few \kms\ \citep[e.g.,][]{kahn54}.  During this
expansion phase, neutral material accumulates between the ionization
front and the shock front that precedes it on the neutral side.  This
massive shell of accumulated neutral material may fragment and give
birth to a second generation of stars -- a process known as the
``collect \& collapse'' process \citep{elmegreen77} of triggered star
formation.  Alternatively, the pressure of the ionized gas can
compress pre-existing clumps and induce their collapse
(``radiation-driven implosion''), again leading to induced star
formation.

If bubbles are two-dimensional structures, star formation triggered by
\hii\ regions would be less prevalent because there would only be a
ring of possible triggering sites instead of a spherical shell.  Using
the ATLASGAL survey at 870\,\micron\ \citep{schuller09} and a sample
of northern bubbles, \citet{deharveng10} found that 38\% of bubbles
are surrounded by collected material and an additional 31\% are
surrounded by several dust condensations.  This suggests that bubble
\hii\ regions may frequently trigger the formation of
second-generation stars, especially massive ones.  \citet{thompson12} estimate that the
  formation of 14\% to 30\% of all massive stars in the Galaxy may be
triggered by the expansion of bubbles.  

There have been numerous studies of young stellar objects (YSOs)
around bubble \hii\ regions \citep[e.g.,][]{deharveng09, watson09,
  beaumont10, deharveng12, samal14}.  These studies have for the most
part identified more YSOs in the direction of the PDRs compared to the number
found in the center of the bubbles.  The statistics for individual
bubble \hii\ regions are sometimes poor, owing to the small number of
YSOs detected per region, but the overall trend is secure.  For
RCW\,120, \citet{deharveng09} found that just two of the $\sim 20$
Class~I and Class~I/Class~II objects are in the direction of the
bubble center.  If bubble \hii\ regions are three-dimensional objects,
we would expect more YSOs in the direction of the bubble centers.

Here, we seek to understand the morphology of IR bubbles and to better
understand their impact in triggering the second generation of stars
by measuring the CO emission and expansion of the angularly large,
nearby bubble RCW\,120 \citep{rcw60}.  Many studies have found
evidence for expansion of molecular material associated with
\hii\ regions \citep{patel95, heyer96, kirsanova08}.  These studies
have largely focused on regions more complicated than RCW120.

\section{The RCW\,120 H{\small II} Region}
RCW\,120 is among the closest \hii\ regions to the Sun, just 1.3\,kpc
distant \citep[see][and references therein]{zavagno07}.  The ionized
gas is $7.5\arcmin$ in diameter, for a physical diameter of $3.8$\,pc.
Optical absorption is seen across the ionized gas and coincident with
weak sub-mm emission, indicating foreground material.  As we show
below, CO is detected for this material at the same velocity as that
of the RCW\,120 bubble, and the absorbing material is therefore local to
RCW\,120.  Because of its nearly spherical appearance in the
mid-infrared, RCW\,120 was dubbed the ``perfect bubble'' by
\citet{deharveng09}.  {\it Herschel} observations of the region
\citep{anderson10a, zavagno10a, anderson12b} show that the PDR of
RCW\,120 emits strongly in the FIR as well.  We show the H$\alpha$
emission from RCW\,120 from SuperCosmos \citep{parker05} in
Figure~\ref{fig:halpha}.  The H$\alpha$ emission from RCW\,120 is
  dominated by the emission from the \hii\ region itself, but also
  shows diffuse emission surrounding RCW\,120.

Star formation is observed everywhere around RCW\,120.  There are
numerous massive condensations along the PDR.  These condensations
were studied in mm-continuum emission by \citet{zavagno07} and by
\citet{deharveng09}.  \citet{anderson12b} give masses derived from
      {\it Herschel} far-infrared data for condensations in the field.
      The most massive condensation, ``Condensation~1'' is $\sim
      800$\,\msun\ and hosts a $\sim300$\,\msun\ source unresolved by
      {\it Herschel}.  \citet{zavagno10a} find that the central source
      in Condensation~1 is a Class~0 YSO of mass 8--10\,\msun\ and an
      envelope mass of $\sim300\,\msun$.  Star formation in the
      fragments along the PDR may have been triggered by the expansion
      of the \hii\ region.  Active star formation
      is also observed in several condensations.  There are numerous
      infrared dark clouds in the field, most of which are covered by
      the Mopra observations of this paper.

\section{Data\label{sec:data}}
The data were taken from 15~July~2011 to 18~July~2011 with the
Australian National Telescope Facility (ATNF) Mopra 22-m radio
telescope near Coonabarabran in New South Wales, Australia.  During
data acquisition the observing conditions were good,
although we lost roughly five hours due to inclement weather.

We used the 3-mm Mopra receiver in wideband mode.  This mode provides
simultaneous spectral coverage of 8.3\,GHz in four spectral windows of
2.2\,GHz, with 8192 channels per spectral window, and dual 
  linear polarizations.  With these tunings we simultaneously
observed the four ground-state ($J = 1\rightarrow 0$) CO isotopologues
$^{12}$CO, $^{13}$CO, C$^{17}$O, and C$^{18}$O at 115.2712\,GHz,
110.2013\,GHz, 112.3590\,GHz, and 109.7822\,GHz, respectively.  (The
$^{13}$CO and C$^{17}$O lines were in the same spectral window.)  We
also tuned to the methanol, CH$_3$OH, masing line $ 0_0
  \rightarrow 1_{-1} E$ at 108.8939\,GHz\footnote{See NRAO's
  Splatalogue for transition frequencies: www.splatalogue.net/}.  Each
tuning provides $\sim 6000$\,\kms\ of spectral range.  The Mopra beam
was $\sim 30\arcsec$ for all tunings, which is $0.19$\,pc at the
distance of RCW\,120.  We corrected the pointing every hour,
  and based on the magnitude of these corrections we estimate the
  pointing accuracy to be better than $10\arcsec$. The spectral
resolution was $\sim0.7$\,\kms\ for all tunings.

We mapped a region $20\arcmin$ square centered at \lb =
(348.282\arcdeg, 0.498\arcdeg) or (J2000 RA, Dec.) = (17:12:22,
  $-$38:25:26).  To create this map we made 16 smaller maps, or
``tiles'', each $5\arcmin$ on a side, in ``On-the-fly'' mapping (OTF)
mode.  We picked one common ``off'' position and observed this
position at the end of each $5\arcmin$ scan row.  The scan-speed was
$2\arcsec$ per second, and the spectra were read out every 2\,s.  We
observed each tile twice, once in the Galactic longitude direction and
once in the Galactic latitude direction, to create a complete map in
each direction.  This method helps greatly to mediate sky noise.  
  Throughout this paper, the units of the molecular data are corrected
  antenna temperature, $T_A^*$.  We corrected for zenith opacity with
  hot and cold loads.

We reduced the data using the ATNF GRIDZILLA \citep{sault95} and
LIVEDATA \citep{barnes01} packages\footnote{http://www.atnf.csiro.au/computing/software/livedata/}.  LIVEDATA takes the raw telescope data, fits
baselines, and prepares the data for gridding into a data cube by
creating SDFITS files.  We fit a linear baseline in LIVEDATA to
emission-free channels for all spectra.  For each position we thus had
four baseline-subtracted spectra: dual polarizations in two scan
directions.  GRIDZILLA takes the SDFITS files created by LIVEDATA and
creates a data cube.  We used GRIDZILLA with a $15\arcsec$ FWHM
Gaussian smoothing to grid the spectra into data cubes, after
averaging the two polarizations.  We set the pixel size to
$12\farcs5$, which is $\sim 40\%$ of the FWHM beam size.  Each data
cube was restricted in velocity from $-100$ to 100\,\kms\ for each
spectral line. The average rms noise per \lb\ pixel location is
0.010\,K, 0.005\,K, 0.006\,K, 0.006\,K,
0.005\,K, respectively for the $^{12}$CO, $^{13}$CO, C$^{17}$O,
C$^{18}$O, and CH$_3$OH lines.

%
%
%

\section{Results\label{sec:results}}
Figure~\ref{fig:total_spectrum} is the spectrum integrated over the
entire map for $^{12}$CO, $^{13}$CO, C$^{17}$O, and C$^{18}$O.  Strong
molecular emission near $-8$\,\kms\ associated with RCW\,120 is
evident from Figure~\ref{fig:total_spectrum}.  As expected, the signal
from $^{12}$CO is the strongest, followed by $^{13}$CO, C$^{18}$O, and
C$^{17}$O.  The peak integrated signal intensity is approximately in the ratio
$1:0.58:0.044:0.0011$ for $^{12}$CO\,:\,$^{13}$CO\,:\,C$^{18}$O\,:\,C$^{17}$O.  We
did not detect CH$_3$OH in these observations and do not analyze the
CH$_3$OH data further.  While C$^{17}$O was detected, the signal is so
weak that we do not provide any additional analysis in the remainder
of this paper.

There are three velocity ranges seen in
Figure~\ref{fig:total_spectrum} that have detected emission.  The
strongest emission is found near $-8$\,\kms.  \citet{zavagno07} found
that the velocity of the ionized gas for RCW\,120 is in the range $-8$
to $-15$\,\kms.  The $-8$\,\kms\ CO emission is therefore associated
with RCW\,120.  There is also faint emission from $-35$ to
$-25$\,\kms, and yet fainter emission from $-75$ to $-50$\,\kms.  In
Figure~\ref{fig:integrated_others} we show $^{12}$CO integrated
intensity maps for the $-35$ to $-25$\,\kms and $-75$ to
$-50$\,\kms\ velocity intervals.  The emission from the velocity
ranges not associated with RCW\,120 has lower intensity, and is also
spatially distinct from the emission associated with RCW\,120.  The
emission from $-35$ to $-15$\,\kms\ seen in
Figure~\ref{fig:integrated_others} is mainly located in the top-left
of the observed region.  This emission is detected in $^{12}$CO,
$^{13}$CO, and C$^{18}$O.  The emission from $-75$ to $-50$\,\kms\ is
filamentary and confined to the bottom of the observed field, but
there are two bright compact sources.  This component is only detected
in $^{12}$CO and $^{13}$CO.  One of these compact sources, near \lb\ =
(348.31\degree, +0.42\degree) is nearly spatially coincident
with a sub-mm clump named ``Condensation~4'' in \citet{zavagno07}.
The other compact source near \lb\ = (348.44\degree, +0.36\degree) has
no obvious FIR or sub-mm counterpart.  In the remainder of this work
we restrict our analysis to the velocity range $-15$ to $+3$\,\kms.

We show the $^{12}$CO, $^{13}$CO, and C$^{18}$O integrated intensity
images for the full $20\arcmin$ map in Figure~\ref{fig:integrated},
integrated over the velocity range $-15$ to +3\,\kms.  The PDR of
RCW\,120 emits most of the CO emission, and the ``interior'' spatially
coincident with the ionized gas is deficient in CO emission relative
to the PDR.  This same basic morphology is seen in the sub-mm
\citep{deharveng09}, mm-wave \citep{zavagno07} and {\it Herschel}
far-infrared regimes \citep{zavagno10a, anderson10a, anderson12b}.
The PDR along the bottom is especially bright in CO, although the PDR
is well-traced over the entirety of the RCW\,120 bubble.

The three isotopologues in Figure~\ref{fig:integrated} show similar
spatial distributions, but there are some notable exceptions.  Most
obvious is the large extended region of emission to the left of the
field that is faint in $^{12}$CO, and bright in $^{13}$CO and
C$^{18}$O.  It is also apparent that the PDR is much thinner in
C$^{18}$O compared to the other isotopologues.  This in part is an
optical depth effect.  As the optical depth increases, we see
self-absorption in $^{12}$CO and $^{13}$CO, but we do not see
self-absorption in C$^{18}$O.  Self-absorption reduces the $^{12}$CO
and $^{13}$CO intensity in the densest regions of the PDR.  The
massive young condensation (``Condensation~1'') along the right
side of the PDR at \lb\ = ($348.18, +0.48$), previously detected at
mm-, sub-mm, and FIR wavelengths, is barely detected in any
isotopologue, indicating strong CO depletion.



We show channel maps of the CO isotopologues in
Figure~\ref{fig:channels}.  These channel maps illustrate the
complicated velocity field of the CO gas, and also show the
differences between the three isotopologues.  In particular, the
region to the left in the fields is very bright in $^{13}$CO and
C$^{18}$O, but not so in $^{12}$CO.  In Figure~\ref{fig:mini} we show
average spectra on top of the integrated intensity images of
Figure~\ref{fig:integrated}.  Each spectrum in Figure~\ref{fig:mini}
is the average of all spectra within the $10\times10$ pixel box.
Multiple velocity components are evident for the $^{12}$CO and
$^{13}$CO data, although for most positions C$^{18}$O only has one
detected velocity component.  A close examination shows that dips in
the \co\ spectra often occur at peaks in the \cor\ and C$^{18}$O
spectra, which is indicative of self-absorption.

We test the optical depth of the \cor\ data using the standard
  LTE analysis.  In LTE, if the excitation
  temperatures ($T_{\rm ex}$) of \co\ and \cor\ are the same, and if
  \co\ is optically thick \citep[see][]{rohlfs},
\begin{equation}
  \tau_0^{13} = -\ln\left[1 - \frac{T_{\rm MB}^{13}}{5.3} \left\{ \left[ e^{5.3 / T_{\rm ex}} - 1 \right]^{-1} - 0.16 \right\}^{-1} \right],
  \label{eq:tau13}
\end{equation}
where $T_{MB}^{13}$ is the main beam brightness temperature for \cor, the 2.7\,K
cosmic microwave background (CMB) is assumed for the background radiation, and
\begin{equation}
  T_{\rm ex} = \frac{5.5}{\ln\left[ 1 + 5.5 / (T_{\rm MB}^{12} + 0.82) \right]}
  \label{eq:tex}
\end{equation}
We assume the CMB here for the background because we estimate the
  free-free emission at 110\,GHz is only $\sim 0.1\,\K$ based on GB6
  data \citep[][assuming optically thin emission]{condon89}, and the
  dust emission is only $\sim 0.03$\,\K based on the {\it Herschel}
  500\,\micron\ data (assuming grey-body emission).

To convert from corrected antenna temperature to main beam brightness
temperature, we assume a main beam efficiency $\eta_{\rm MB} = 0.40$
\citep[see][]{ladd05}: $T_A^* = \eta_{\rm MB} T_{\rm MB}$. Using the
channel with the greatest peak intensity ($-7.17\,\kms$ for \co\ and
$-7.46\,\kms$ for \cor), Equation~\ref{eq:tex} yields average
excitation temperatures over the entire map of $T_{\rm ex} =
14.6\pm3.6$\,K.  Using only the 50 pixels with the highest
\cor\ intensity at $-7.46\,\kms$, the average value is $T_{\rm ex} =
23.4\pm4.8$\,K.  We find that 64\% of all $12\farcs5$ map pixels
at $-7.46\,\kms$ have \cor\ optical depths less than unity.  Nearly
all of the 50 brightest pixels at $-7.46\,\kms$, however, have optical
depths greater than unity, with an average value $>2$.  The average
here is approximate because a fraction of \cor\ pixels are quite
optically thick, in which case Equation~\ref{eq:tau13} breaks down.
In the central $20\times20$ pixels, the average \cor\ optical depth is
$1.05\pm0.69$ at $-7.46\,\kms$.

We repeated the calculations to evaluate the C$^{18}$O optical depth
(the form and constants in Equation~\ref{eq:tau13} are unchanged for
C$^{18}$O) and found that all pixels have optical depths of C$^{18}$O
less than unity for the $-7.40\,\kms$ channel.  We conclude that
\cor\ is for the most part optically thin, although it is optically
thick along the PDR for the highest intensity channel, while C$^{18}$O
is optically thin everywhere.


\section{Tests for expansion of the molecular gas}
The goal of these Mopra CO observations is to test for expansion of
the RCW\,120 bubble.  For three-dimensional objects, expansion along
the line of sight will result in blue-shifted emission from the
near-side of the object, and red-shifted emission from the far side.
These wavelength shifts are in relation to the bubble itself
  (i.e., the PDR), which will be at a velocity in between that of the
near and far side material.  If the red- and blue-shifted velocity
components cannot be spectrally resolved, the line width will be
broader where the expansion is toward and away from the observer.  

The optical image of RCW\,120 clearly shows absorption from foreground
material, running across the face of RCW\,120.  If RCW\,120 is
carrying this material through its expansion, we would expect that its
spectrum would be blue-shifted relative to the bulk of the emission in
the region.  This absorption feature provides an ideal near-side
filament to test for blue-shifted emission. While it is possible
  that the absorbing filament is not part of the PDR, but rather is
  part of a quiescent molecular cloud foreground to the PDR, we regard
  this option as unlikely.  The absorption caused by the filament, seen in
  Figure~\ref{fig:halpha}, stops at the ionized gas boundary.  If the
  filament were part of a larger molecular envelope, we would expect
  its absorption to be spread over a larger angle, extending beyond
  the face-on PDR o RCW120.


In the following, we use only $^{13}$CO, and C$^{18}$O data.  We
exclude the $^{12}$CO data because optical depth effects may confuse
the analysis.


\subsection{Spectra}
The most obvious sign of expansion would be the detection of two
velocity components in the center of the bubble, one blue-shifted and
one red-shifted with respect to the velocity of the PDR gas.  If
  the expansion speed is slower than the line width, the expansion
  signal may appear as enhancements in the line wings.  A high optical
  depth will complicate the interpretation of the spectra.

Most $^{13}$CO spectra in Figure~\ref{fig:mini} toward the interior of
RCW\,120 have two velocity components.  The situation is
complicated by the fact that two components are also seen toward the
PDR in many locations, and even outside the bubble in the bottom left of
the maps.  Therefore, it is difficult to draw firm conclusions from Figure~\ref{fig:mini}.

The dark absorption across the face of RCW\,120 offers another way to
test for expansion.  We expect that this material is on the near-side
of RCW\,120, and that its associated molecular gas will be
blue-shifted relative to the PDR of RCW\,120.  We test this hypothesis by
extracting spectra averaged over an aperture defined visually to
approximate the extent of the absorbing gas (hereafter the
``Absorption'' aperture).  We also create average spectra from the
entire interior of the bubble (``Interior'' aperture), at locations of
the interior of the bubble where the optical absorption is not seen
(``No Absorption'' aperture), and from the PDR (``PDR'' aperture).
These apertures are shown in Figure~\ref{fig:apertures}.  If there is
three-dimensional expansion, we expect that the spectrum from the
Absorption aperture has blue-shifted emission, that the spectrum from
the No Absorption aperture has red-shifted emission, that the spectrum
from the Interior aperture has a mix of these two qualities, and that
the velocity from the PDR aperture is in between that of any red- or
blue-shifted gas.

We show the spectra created from the apertures defined above in
Figure~\ref{fig:interior}.  All spectra are normalized by their peak.
We find little difference between the four different apertures.  For
the $^{13}$CO data, all apertures can be modeled with two velocity
components, one near $-10$\,\kms\ and one near $-8$\,\kms.  Compared
to the other apertures, the $-10$\,\kms\ component of the Absorption
aperture is fainter relative to the $-7\,\kms$ component.  For
C$^{18}$O, again the Absorption aperture spectrum shows decreased
intensity of the $-10$\,\kms\ component relative to the $-7\,\kms$
component.  We note that the $-10$\,\kms\ component is quite weak for
many of the C$^{18}$O aperture spectra.  If the Absorption aperture
  has more blue-shifted gas, we would expect the
  $-10$\,\kms\ component to be brighter for this aperture, not
  fainter.  Because of the similarity in the shape of the spectra from
  the four apertures, we conclude that these spectra show no
  discernible expansion relative to the PDR.

In contrast to the Trifid nebula, another bubble \hii\ region,
  the material across the face of RCW\,120 is not detected in emission at
  8.0\,\micron.  For Trifid, the dark dust lanes seen in optical
  images appear bright at 8.0\,\micron\ \citep[][their
    Figure~1]{rho06}.  For RCW\,120, the dust lane is not seen in
  either absorption or emission at 8.0\,\micron.
  Figure~\ref{fig:interior} clearly shows that the dust lane is
  associated with RCW\,120 in velocity, and the fact that it is
  observed in absorption in H$\alpha$ places it in front of the
  ionized gas.  The filament is seen in emission at
  870\,\micron\ \citep[][their Figure~2]{deharveng09}, indicating that
  it is cold.  The 870\,\micron\ emission, however, is faint compared
  to the emission from the PDR, consistent with our CO observations.
  We hypothesize that the filaments in the Trifid are more deeply
  embedded in the \hii\ region, causing the emission at 8.0\,\micron,
  and furthermore that they are denser than those of RCW\,120.  This
  may be why the filament in RCW\,120 is not detected in emission at
  8.0\,\micron\ (not embedded in the ionized gas) or absorption (too
  diffuse to cause mid-infrared emission).


\subsection{Channel Maps}
Something that is difficult to see in the average spectra can be
easier to see in channel maps.  Again, if expansion is detected, we
expect to see blue- and red-shifted material in the interior of the
bubble.  In the channel maps of Figure~\ref{fig:channels}, this would
manifest itself as emission in the interior at low velocities (left
side of figure), emission from the PDR at the source velocity (middle
of figure), and emission from the interior again at high velocities
(right side of figure).  This trend is not seen.  For C$^{18}$O, in
fact, we see that the ``Absorption'' aperture region is brightest in
the same channel where the PDR emission is the brightest.  This
indicates that there is little if any expansion relative to the PDR.

\subsection{Second Moment Maps}
It is possible that our spectral resolution is not sufficient to
detect expansion in the above methods. We therefore create maps of the
intensity-weighted velocity dispersion, known as second moment maps
(Figure~\ref{fig:moment2}).  For expanding three-dimensional bubbles,
the velocity dispersion would increase in the bubble interior relative
to that toward the PDR.  The second moment maps for $^{13}$CO and
C$^{18}$O do not show such a trend, and in fact the velocity
dispersion along the PDR is in many places larger than that of the
bubble interior.  The average intensity-weighed C$^{18}$O
  velocity dispersion in the direction of the interior is $0.88\pm
  0.46$\,\kms, and it is $0.80\pm0.29$\,\kms\ for the PDR (measured in
  a $0.1\degree$ exterior radius circular annulus that excludes the
  interior region).

These molecular data show a slightly different picture from the one of
\citet[][their Figure~10]{zavagno07}.  Their H$\alpha$ map of RCW\,120
has a line width in the interior of $\sim25$\,\kms, and a line width
toward the ionization front of $\sim21$\,\kms.  Assuming a uniform
electron temperature across the entire region of 7100\,K (see
Section~\ref{sec:discussion}), this would indicate that the thermal
line width is $\sim 15$\,\kms.  Adding the thermal and non-thermal
line widths in quadrature, we find a non-thermal line width of
20\,\kms\ for the interior and 15\,\kms\ near the ionization front.
Both turbulence and ordered expansion contribute to the non-thermal
line width.  The comparison between the molecular and ionized gas is
not straightforward though.  The H$\alpha$ observations probe the
entire volume of ionized gas while the CO observations mainly probe the
shell of collected material.  For larger scales (i.e., the line of
sight in the direction of the bubble interior), we would expect
increased line widths, as is seen for molecular clouds
\citep[e.g.][]{larson81}. We therefore cannot take the increased
H$\alpha$ line widths in \citet{zavagno07} as evidence for expansion.

\subsection{Position-Velocity Diagrams}
Finally, we make position-velocity diagrams (Figure~\ref{fig:pv}) for
\cor\ and C$^{18}$O by integrating the cube over one spatial
direction.  Expansion in a shell should create a circular structure in
such a figure.  There are a few interesting features in these
diagrams, but nothing that can be directly atributed to expansion.

First, there is blue-shifted material near $-12$\,\kms\ evident
at \lb = ($348.28\degree, +0.45\degree$), and \lb = $(348.28\degree,
+0.35\degree)$.  This emission is evident in both isotopologues,
although it is much more obvious in \cor.  This emission is from a
cloud found toward the middle of the $-12$\,\kms\ (leftmost) panels in
the channel maps (Figure~\ref{fig:twovel}, see following discussion of
this feature).  This cloud extends toward the bottom of the maps,
beyond the PDR, and therefore does not indicate expansion.  It is
largely the emission from this cloud that appears as the second
component in the aperture photometry spectra
(Figure~\ref{fig:interior}).  It is therefore not too surprising that
the spectra from the various apertures do not show much difference.

The brightest emission in the position-velocity diagrams, and
especially that of the C$^{18}$O longitude-velocity diagram, has a
curved shape. Because the PDR extends across the entire frame, this
does not indicate that the interior gas is necessarily red-shifted
with respect to that of the PDR.  Rather, nearly all the
emission in Figure~\ref{fig:pv} is from the PDR, because the interior
emission is rather faint by comparison.

We illustrate these two features in Figure~\ref{fig:twovel}, which
shows \cor\ emission at $-12$\,\kms\ (left panel), $-6$\,\kms\ (right
panel) on top of GLIMPSE 8.0\,\micron\ emission mainly from the PDR.
The $-12$\,\kms\ emission is seen toward the left of the field, and
outside the PDR on the bottom of the field.  The $-6$\,\kms\ emission
is strongest outside the PDR on the left and right sides of the field.
It is this $-6$\,\kms\ emission that is causing the curved shape in
the longitude-velocity diagrams.

\section{Discussion\label{sec:discussion}}
We find no conclusive evidence of expansion in the Mopra molecular
data.  This suggests that RCW\,120 is a two-dimensional structure, or
that it is spherical but expanding at a rate lower than that
detectable in our data.  The presence of absorption across the face of
RCW\,120 from material associated with the region, however, implies
that the two-dimensional geometry cannot be correct.  It is also
possible that RCW\,120 is only half a bubble, which would reduce our
estimated expansion by a factor of two.  Below, we assess the
theoretical expansion rate of RCW\,120, assuming spherical symmetry.

\citet{spitzer78} derived an expression for the time evolution of an \hii\ region:
\begin{equation}
R(t) = R_{S}\left(1 + \frac{7c_it}{4 R_{S}}\right)^{4/7}\,,
\label{eq:r_evolve}
\end{equation}
where $R$ is the \hii\ region radius, $R_{S}$ is the initial Str{\"o}mgren
radius, $c_i$ is the sound speed in the ionized material, and $t$ is the
elapsed time since the Str{\"o}mgren radius formed, in seconds.  
Therefore,
\begin{equation}
  \frac{dR}{dt} = c_i \left( 1 + \frac{7c_it}{4 R_{S}}\right)^{-3/7} = c_i \left(\frac{R_S}{R}\right)^{3/4}\,.
\label{eq:spitzer_expand}
\end{equation}
The Str{\"o}mgren radius, the initial equilibrium size between ionization and recombination, is:
\begin{equation}
R_{S}=\left( \frac{3}{4 \pi} \frac{N_{\rm Lyc}}{n_0^2 \beta_2} \right)^{1/3}\,,
\label{eq:stromgren}
\end{equation}
where $n_0$ is the density of the ambient medium around the \hii\ region, and $\beta_2$
is the recombination coefficient, in units of ${\rm cm^3\,s^{-1}}$.
Due to a pressure imbalance between the \hii\ region plasma and the
neutral material, the expansion of an \hii\ region does not stop at
the Str{\"o}mgren radius.  
%
%
%
%


The initial Str{\"o}mgren radius depends on $N_{\rm Lyc}$ and $n_0$.
\citet{martins10} found that RCW\,120 is ionized by a single star of
spectral type O6-8V/III.  These spectral types have a range of
ionizing photon rates from $10^{48.29}$\,s$^{-1}$ for an O8V star to
$10^{49.22}$\,s$^{-1}$ for an O6III star \citep{martins05}.
\citet{kuchar97} find an integrated radio flux of 5.5\,Jy at 5\,GHz
for RCW\,120, and therefore $N_{\rm Lyc} = 10^{47.85}$\,s$^{-1}$ for
$T_e=10^4$\,K using the equation from \citet{rubin68}.
\citet{caswell87} find 8.3\,Jy at 5\,GHz, which leads to $N_{\rm Lyc}
\simeq 10^{48.0}$\,s$^{-1}$.  Even if half the ionizing photons escape
or are absorbed by dust within RCW\,120, and for a decreased electron
temperature of 5000\,K, we still only find $N_{\rm Lyc} \simeq
10^{48.5}$\,s$^{-1}$.  We therefore think the high end of the ionizing
photon ranges quoted above is unrealistic, and use $N_{\rm Lyc} =
10^{48.29}$\,s$^{-1}$ to $10^{48.63}$\,s$^{-1}$ in the following
  calculations, which corresponds to O7V-O8V stars \citep{martins05}.

The ambient density, $n_0$ is difficult to estimate.  From Larson's
laws \citep{larson81}, for molecular clouds $n = 3400 L^{-1.10}$,
where $n$ is the molecular hydrogen density in \percc and the cloud
size $L$ is in pc.  For a radius of 1.9\,pc, the density $n$ is
  $\sim 1700\,\percc$.  We are only concerned with the number of
atoms that can be ionized, so this would imply that $n_0 =
3400$\,\percc.  Using {\it Herschel} data and a dust-to-gas mass of
100, \citet{anderson12b} found 2000\,\msun for the entire RCW\,120
bubble.  Averaged over a sphere 1.9\,pc in radius, the mean atomic
density is 2200\,\percc.  Assuming all this material was swept up
during the expansion of RCW\,120, 2200\,\percc\ is the mean density of
the natal cloud.  These are both lower limits to the true density.  In
both cases we do not know if the original cloud extended to the
current radius of RCW\,120.  Furthermore, there must have been a
density profile such that the center of the cloud would have had a
higher density.  We therefore take 2000\,\percc\ to 6000\,\percc\ to
be a reasonable range for $n_0$.


For the assumed range of ionization rates and ambient densities, we
find $R_S \simeq 0.12$ to 0.32\,pc.  This calculation depends on the
electron temperature of the plasma.  \citet{caswell87} measured the
radio recombination line emission and radio continuum emission from
RCW\,120 at 5\,GHz.  Using their measured line and continuum
intensities, we find an electron temperature of 7100\,K, assuming the
helium to hydrogen abundance ratio of 0.08 \citep{quireza06a}, and
using the equations in \citet{balser95} and \citet{quireza06b}.  To
calculate $R_S$, we used $\beta_2 = 2 \times 10^{-10} T_e^{-3/4}$
\citep{dyson}, which yeilds $\beta_2 = 2.59 \times
10^{-13}\,\percc$\,s$^{-1}$.  The current radius of RCW\,120 is
therefore 7--16\,$R_S$.  From Equation~\ref{eq:spitzer_expand}, we
expect RCW\,120 to be expanding at 1.3 to 2.3\,\kms\ \citep[see
    also the model in][]{zavagno07}.  This assumes $c_i\simeq
10$\,\kms, which is appropriate for $T_e = 7100$\,K.


Taking into account the ambient pressure surrounding the \hii\ region,
\citet{raga12} find for the expansion rate of \hii\ regions:
\begin{equation}
  \frac{1}{c_i} \frac{{\rm d} R}{{\rm d} t} = \left( \frac{R_{S}}{R} \right) ^{3/4} - \sigma \left( \frac{R}{R_{S}} \right) ^{3/4},
\end{equation}
where $R$ is the \hii\ region radius, and $\sigma = c^2_n / c^2_i$,
the squared ratio of the ionized to ambient sound speeds.  This
reduces to Equation~\ref{eq:spitzer_expand} if the final term is
removed.  If $c_n \simeq 0.3$ to 1.1\,\kms\ (corresponding to
molecular hydrogen at a temperature of 15\,K and neutral hydrogen of
100\,K), $\sigma \simeq 0.01$ to $0.001$.
According to this model, RCW\,120 should still be expanding at 1.2 to
2.3\,\kms (essentially unchanged from above calculations).  These
values are consistent with our lack of detected expansion (see below).

Tremblin et al. (2014, submitted) create models that take into account
the turbulence of the ambient gas.  For RCW\,120, they find an
expansion velocity of $\sim1.6$\,\kms\ for the ionization front.
This is again roughly consistent with our lack of detected expansion.
Furthermore, they show that magnetic fields and gravity for a density
profile $\rho\propto r^{-\alpha}$, with $\alpha > 0.5$, have a negligible
effect compared to the turbulence.  In a homogeneous medium, the
expansion could be slowed further due to the gravity of the central
cluster.  For RCW\,120, 200 to $600\msun$ is required for
$n_0\simeq2000$ to 6000\percc.  While such a cluster mass is possible for a
normal initial mass function and a 30\,\msun\ ionizing source,
no cluster has yet been detected in RCW\,120.  There is thus
no evidence to suspect that self-gravity is important in slowing
the expansion of the region.

How can we interpret these results?  All models are able to reproduce
the lack of detected expansion.  Detecting the expansion of a
classical \hii\ region is difficult with such molecular data. This
unfortunately implies that we cannot test for the
three-dimensional nature of RCW\,120 using these data.

\subsection{Simulated C$^{18}$O Spectra}
To better understand our data, we model the expansion of a 1.9\,pc
radius molecular shell of thickness 0.4\,pc (20\% of the radius),
using the three-dimensional line radiative transfer code
\textsc{MOLLIE}
\citep{keto04,rawlings04,redman06,carolan08,lo11}. \textsc{MOLLIE} has
been benchmarked against a suite of test problems as per
\citet{vanzadelhoff02}, and the models were found to reproduce the
test observations to within a few per cent.  In order to calculate the
level populations, the statistical equilibrium equations are solved
using an accelerated lambda iteration \citep{rybicki91}, which reduces
the radiative transfer equations to a series of linear problems that
are solved quickly even in optically thick conditions.  For an
arbitrary viewing angle to the model cube, ray-tracing is then used to
calculate the molecular line intensity as a function of velocity, for
a set of positions that matches in number the fixed gridding of the
model.

The input to \textsc{MOLLIE} is divided into voxels and there are five
input parameters that need to be defined for each voxel: the gas bulk
velocity, the gas turbulent velocity, the relative abundance of the
molecular species of interest, the gas temperature, and the number
density of H$_2$.  We create two models, with expansion velocities of
1.0\,\kms\ and 1.5\,\kms.  We use a value of 1.5\,\kms\ for the
magnitude of the turbulent velocity (approximated by inspection of
individual \cor\ spectra), an abundance n(\cor)/n(H$_2$)$ = 5 \times
10^{-7}$ \citep[from][]{stahl08, lee96}, a temperature of 30\K,
and a H$_2$ density of $1\times10^4$\,\percc.  The temperature and
H$_2$ density are derived from the dust temperature and column density
values from \citet{anderson12b}, for our assumed geometry (the
temperature is also broadly consistent with the \co\ intensity, with
an assumed main beam efficiency of 0.4).  Our model grid has
$16\times16$ spatial locations.

From our model, we create synthetic \cor\ spectra, shown in
Figure~\ref{fig:model}.  The ``central'' spectrum is from a single
pixel at the center of the grid, while the ``PDR'' spectrum is from a
single pixel along the expanding shell boundary.  We see that there is
no double-peaked expansion profile detectable for 1\,\kms, but there
is for 1.5\,\kms.  Compared with the \cor\ data in
Figure~\ref{fig:apertures}, we see that our data are more consistent
with an expansion speed of 1.0\,\kms.  A turbulent velocity that
exceeds the bulk velocity significantly will always mask an expansion
signature, but this turbulent velocity is fairly well-constrained by
our data.  Our model suggests that RCW\,120 is expanding slower than
1.5\,\kms.


\section{Holes in the perfect bubble}
Deficits in the CO emission along the PDR of RCW\,120 are apparent in
Figure~\ref{fig:channels}.  Assuming these emission deficits are true
holes in the PDR, they will allow radiation from the central star(s)
to escape, ionizing and heating the ISM.  \citet{anderson10a} noted
numerous places beyond the ionization front of RCW\,120 where the dust
temperature, calculated from {\it Herschel} data, was enhanced, most
notably along the bottom PDR.  This indicated that photons from RCW
were leaking into the ISM.  Because the PDR of RCW\,120 appears
relatively uniform in the {\it Herschel} far-infrared data, it was
slightly unclear if this interpretation was correct.

In Figure~\ref{fig:enhancements} we attempt to identify these
temperature enhancements outside the PDR more clearly.  Here we show
the GLIMPSE 8.0\,\micron\ data with a single contour of {\it Herschel}
dust temperature from \citet{anderson12b}, at 22\,K.  This
  temperature is an average of that found interior to the PDR (for
  which \citet{anderson12b} found 24.9\,K) and the background
  interstellar medium (for which \citet{anderson12b} found
  $\sim19$\,K).  We mark significant dust temperature enhancements
outside the PDR with lines.  These enhancements are located along the
southern and western edges of the bubble, and coincide with
discontinuities in the PDR seen at 8.0\,\micron\ \citep[noted
  previously in ][]{zavagno07}.  The magnitude of the enhancements is
$\gsim2\,K$.  The temperature map did not include 8.0\,\micron\ data,
and therefore the fact that the temperature enhancements outside the
PDR correlate with locations where the PDR is less well-defined at
8.0\,\micron\ is not a coincidence.

We see in Figure~\ref{fig:holes} that the enhanced dust temperatures
are spatially correlated with the deficits in CO emission, and that
the CO deficits are also correlated with discontinuities in the PDR
observed at 8.0\,\micron, and H$\alpha$ emission outside the PDR.  The
green lines indicate positions outside the PDR that have enhanced dust
temperatures, from Figure~\ref{fig:enhancements}
\citep[cf.][]{anderson10a}.  The dust temperature enhancements to the
west and east coincide with a lack of CO in the $-8.9$\,\kms\ channel,
while those to the south are between CO clumps seen in the
$-6.0$\,\kms\ channel.  All evidence suggests that these
temperature enhancements are real, caused by the porous PDR of
RCW\,120, and that the holes in the PDR can be associated with
particular CO velocities.

\begin{figure} \centering

\includegraphics[scale=0.80]{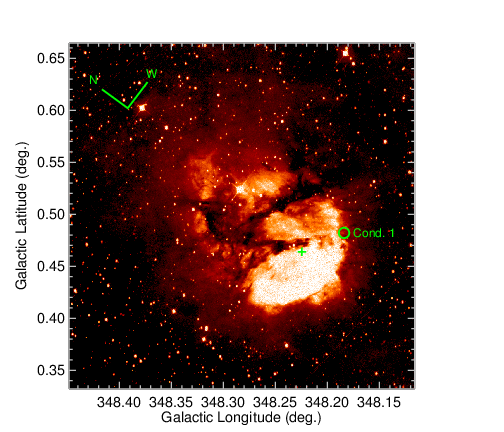}
\caption{H$\alpha$ emission of RCW\,120 from SuperCosmos
  \citep{parker05}.  The map covers the same area as our Mopra
  observations, and is $20\arcmin$ square, centered at \lb =
  (348.282\arcdeg, 0.498\arcdeg), and oriented in Galactic
  coordinates.  The north and west directions are indicated.  The
  location of the ionizing source \citep[from][]{martins10} is
  identified with a small green cross and the location of
  ``Condensation~1'' is identified with a green circle.  Absorption
  across the face of the \hii\ region is clearly shown, as is diffuse
  extended H$\alpha$ emission outside the main ionized zone.}
\label{fig:halpha}
\end{figure}

\begin{figure} \centering

\includegraphics[scale=0.80]{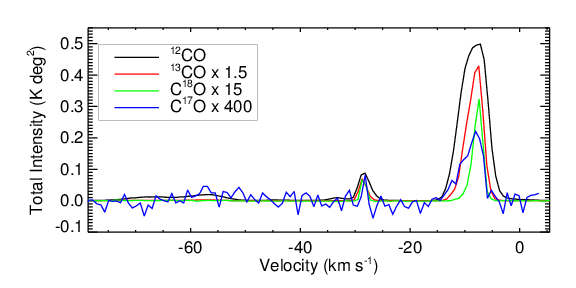}
\caption{Average spectra (in $T_A^*$) integrated over the entire map.
  The emission from RCW\,120 near $-7$\,\kms\ is evident in all CO
  isotopologues.  There is also emission near $-30$\,\kms\ and
  $-60$\,\kms that is not associated with RCW\,120.}

\label{fig:total_spectrum}
\end{figure}

\begin{figure} \centering

\includegraphics[height=3in]{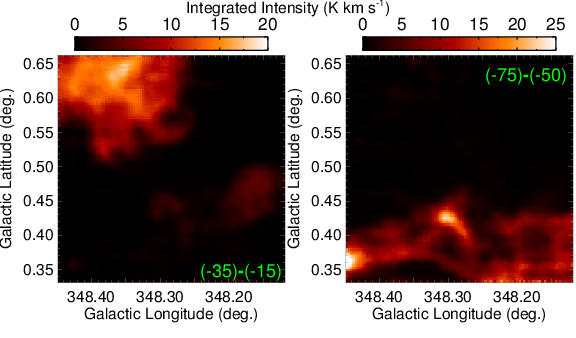}

\caption{Integrated intensity images of $^{12}$CO for two velocity
  ranges not associated with RCW\,120 (see also
  Figure~\ref{fig:total_spectrum}).  The velocity ranges of
  integration are $-35$ to $-15$\,\kms\ (left panel) and $-75$ to
  $-50$\,\kms\ (right panel), as shown on the maps.}

\label{fig:integrated_others}
\end{figure}

\begin{figure} \centering

\includegraphics[width=6.5in]{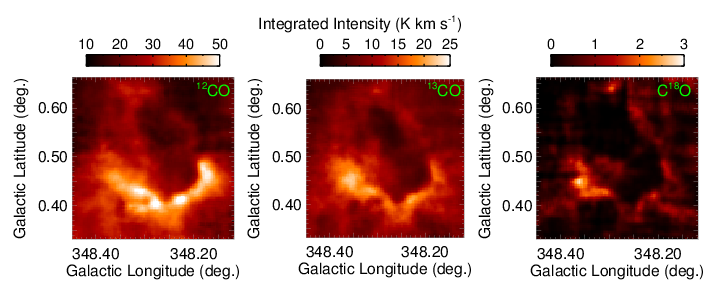}

\caption{Integrated intensity images of CO isotopologues.  All data
  are integrated from $-15$ to +3\,\kms\ and smoothed with a Gaussian
  filter of FWHM 3 pixels.  The PDR is the brightest feature, and it
  is especially prominent to the bottom of the field.  The brightest
  location of C$^{18}$O emission is to the left in the field, near \lb
  = (348.36, 0.45), where $\cor$ is also the brightest; this location
  is relatively faint in $^{12}$CO.}

\label{fig:integrated}
\end{figure}

H$\alpha$ emission beyond the ionization front of RCW\,120 was noted
by \citet{zavagno07} and \citet{deharveng09}, who attributed it to the
porous PDR.  This emission can be clearly seen in
Figure~\ref{fig:halpha}.  There are two components of this emission:
bright discrete regions just beyond the ionzation front (most evident
for the bottom PDR) and a low intensity ``halo'' surrounding RCW\,120
on all sides.  This halo is not perfectly symmetric, but rather is
more extended toward the top of the observed field.  The bubble itself
appears open toward the top of the field in the $-6.0$\,\kms\ channel
seen in Figure~\ref{fig:holes}.  The increased diffuse H$\alpha$
intensity toward the top of the field is likely due to the fact that
the PDR is weak or absent in this direction at some velocities.  The
locations of the bright discrete regions outside the PDR coincides
with the temperature enhancements in Figure~\ref{fig:holes}, and with
a lack of CO emission.


We estimate the amount of flux leaving RCW\,120 through these holes
using the SuperCosmos H$\alpha$ data of Figure~\ref{fig:halpha}.
Because of the absorption across the face of this region, the
H$\alpha$ emission underestimates the true H$\alpha$ flux.  We crudely
account for this by assuming that the emission in the absorbing zones
across the face of RCW\,120 is of the same intensity as that of nearby
zones that appear to lack absorption.  We perform aperture photometry
on the entire region within the PDR (including the absorbing zones
``filled in'' with nearby average values), and individually on each of
the zones outside the PDR identified with green lines in
Figure~\ref{fig:holes}, with a local background that is representative
of the diffuse H$\alpha$ field just outside the PDR.  About $5\%$ of
the total intensity is emitted in the zones outside the nominal
ionization front, identified with green lines in
Figure~\ref{fig:holes}.  We stress that because of the absorption
across RCW\,120, this number is rather uncertain.

RCW\,120 has a diffuse H$\alpha$ ``halo,'' most likely due to photons
leaking from smaller holes in its PDR.  Performing aperture photometry
over the larger RCW\,120 region with an aperture $15\arcmin$ in
diameter, we find that $\sim25\pm10$\% of the intensity associated
with RCW\,120 is from outside of its PDR.  The uncertainties come from
different background choices in the aperture photometry.  This figure
includes the 5\% for the discrete zones.

\section{Conclusions}
We observed the bubble \hii\ region RCW\,120 with the Mopra radio
telescope in the {$J = 1\rightarrow 0$ transitions of} \co, \cor,
C$^{18}$O, and C$^{17}$O.  We also observed, but did not detect
emission from the 108.8939\,GHz methanol maser transition.  The goal of
these observations was to search for expansion of the molecular
material in the foreground and background PDRs of RCW\,120.  Such
expansion would indicate that RCW\,120 is a three-dimensional
structure.  We do not find conclusive evidence for expansion.  This
result is in rough agreement with models for the time-evolution of the
RCW\,120 \hii\ region, and we therefore can make no claims on the
three-dimensional nature of this \hii\ region.

We note that CO observations from a single-pointing would not
necessarily reach this same conclusion.  Many spectra toward RCW\,120,
have multiple velocity components in optically thin tracers, and this
could be taken as evidence for expansion.  This should be accounted
for in future, similar studies.

Although the strongest CO emission is along the PDR, there are
numerous ``holes'' in the PDR CO emission.  The locations of these
holes correspond to dust temperature enhancements seen in {\it Herschel}
data \citep{anderson10a, anderson12b}.  This indicates that photons
are leaving the nearly complete PDR of RCW\,120 in discrete locations.
An H$\alpha$ map of the region confirms this interpretation.  We find
that $25\pm10\%$ of the H$\alpha$ flux is found outside the PDR of
RCW\,120, with $\sim5\%$ found at the locations of these discrete
holes.

\begin{acknowledgments}
We thank the referee for a thorough reading, and for insightful
comments that improved the clarity of this manuscript.  L.D.A thanks
the staff of the Laboratoire d'Astrophysique de Marseille for their
hospitality and friendship during final preparation of this
manuscript.  The Mopra radio telescope is part of the Australia
Telescope which is funded by the Commonwealth of Australia for
operation as a National Facility managed by CSIRO.
L.D.A. acknowledges support by the ANR Agence Nationale for the research
project ``PROBeS'', number ANR-08-BLAN-0241.
\end{acknowledgments}

\bibliographystyle{apj} 
\bibliography{ref.bib} 






\clearpage


\clearpage

\clearpage

\clearpage

{\rotate
\begin{figure} \centering
\includegraphics[width=6.5in]{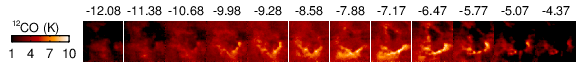}
\includegraphics[width=6.5in]{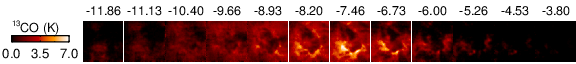}
\includegraphics[width=6.5in]{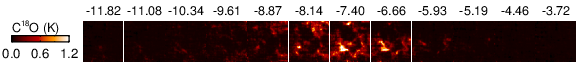}

\caption{Channel maps (in $T_A^*$) of $^{12}$CO (top), $^{13}$CO
  (middle), and C$^{18}$O (bottom).  All images of individual
    channels have the same size as Figures~1, 4, and 5: $20\arcmin$
    square, centered at \lb = (348.282\arcdeg, 0.498\arcdeg), and
    oriented in Galactic coordinates. The velocity in \kms\ is given
    above each image.  All panels of a given isotopologue share the
  same intensity scaling.  The velocity ranges between isotopologues
  differ slightly due to the different spectral resolutions.}

\label{fig:channels}
\end{figure}
}
\clearpage


\begin{figure} \centering

\includegraphics[width=2in]{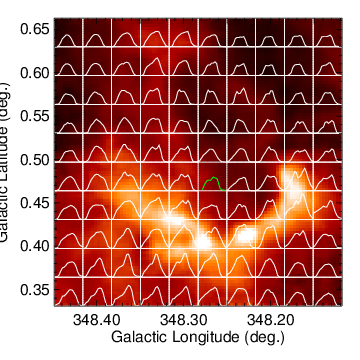}
\includegraphics[width=2in]{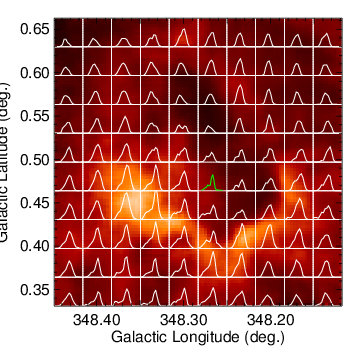}
\includegraphics[width=2in]{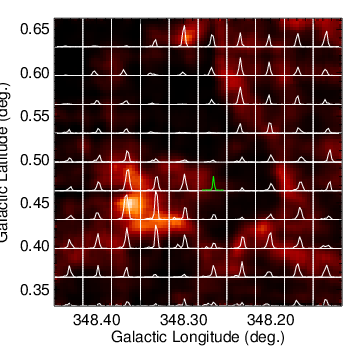}
\includegraphics[width=2in]{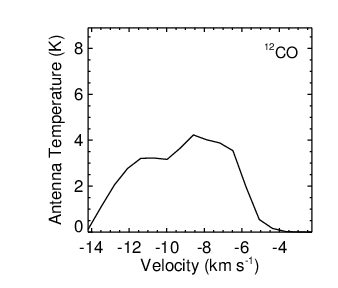}
\includegraphics[width=2in]{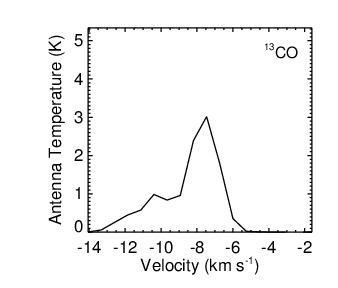}
\includegraphics[width=2in]{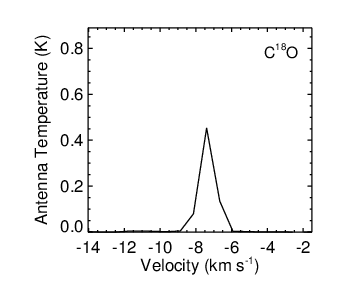}

\caption{(Top row) Spectra overlaid in the integrated intensity images
  of $^{12}$CO (left), $^{13}$CO (middle) and C$^{18}$O (right).  Each
  small box contains 100 spectra in the respective data cubes, and the
  spectra shown are the average of all spectra within the box.
  (Bottom row) $^{12}$CO (left), $^{13}$CO (middle) and C$^{18}$O
  (right) average spectra from the location where spectra in the top
  row are shown in green.  The scaling and velocity range of all
  individual spectra in the top images is the same as those shown in
  the bottom row.}

\label{fig:mini}
\end{figure}
\clearpage

\begin{figure} \centering

\includegraphics[width=3in]{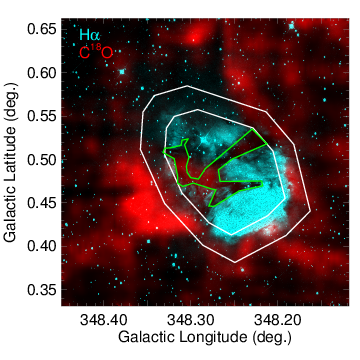}
\includegraphics[width=3in]{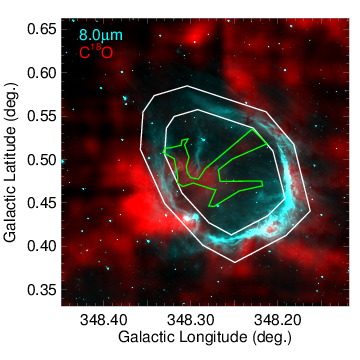}
\caption{{\it Left:} Apertures on top of C$^{18}$O (red) and
    SuperCosmos H$\alpha$ (blue) images of RCW\,120.  {\it Right:} The
    same, for GLIMPSE 8.0\,\micron\ data in blue.  We call the green
    aperture ``Absorption'', the interior white aperture ``Interior,''
    the space between the two white apertures ``PDR,'' and the space
    between the Absorption and Interior apertures ``No Absorption.''
    There is excellent morphological agreement between the absorption
    seen in the H$\alpha$ data and the C$^{18}$O emission across the
    face of RCW\,120 (enclosed by the Absorption aperture).  The
    GLIMPSE 8.0\,\micron\ data show that the PDR is enclosed by our
    PDR aperture.}

\label{fig:apertures}
\end{figure}
\clearpage

\begin{figure} \centering

\includegraphics[width=5in]{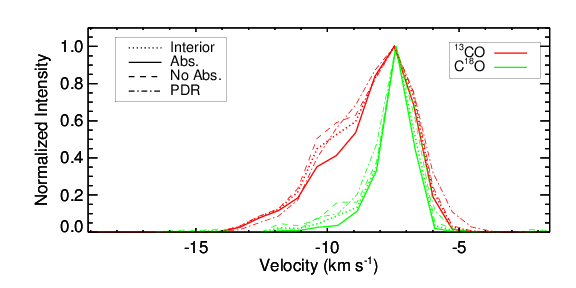}
\caption{Spectra from the four apertures defined in the text and in Figure~\ref{fig:apertures}.  So the
  line profiles can be better seen, we normalized all spectra to a
  peak corrected antenna temperature, $T_A^*$, of 1\,K.}

\label{fig:interior}
\end{figure}
\clearpage

\begin{figure} \centering

\includegraphics[width=5in]{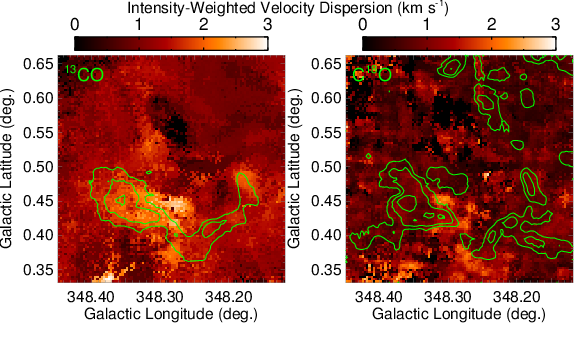}
\caption{Second moment (intensity-weighted velocity dispersion) maps
  of \cor\ (left) and C$^{18}$O (right).  Contours show integrated
  intensity values from Figure~\ref{fig:integrated}, and have values of 14, 18,
  and 22\,K\,\kms\ for \cor\ and 0.7, 1.2, and 2\,K\,\kms\ for
  C$^{18}$O.  The velocity dispersion is not higher toward the
  interior of RCW\,120, and therefore there is no signature of
  expansion in the second moment maps.}
\label{fig:moment2}
\end{figure}
\clearpage


\begin{figure} \centering

\includegraphics[scale=0.70]{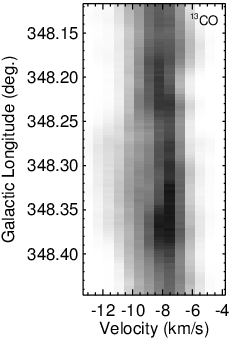}
\includegraphics[scale=0.70]{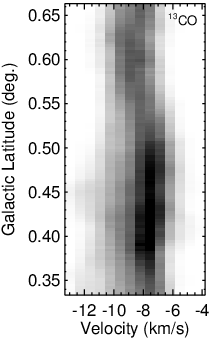}
\includegraphics[scale=0.70]{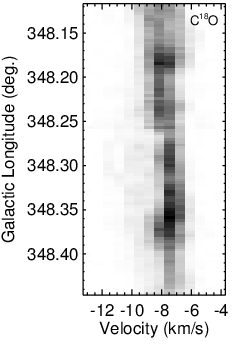}
\includegraphics[scale=0.70]{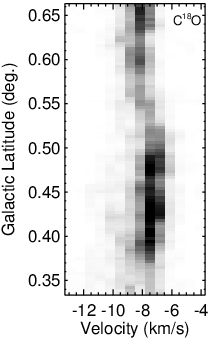}
\caption{Position-velocity diagrams created by integrating the data
  cubes along one spatial direction.  The top row shows \cor\ and the
  bottom row shows C$^{18}$O.  In the left column are
  longitude-velocity diagrams and in the right column are
  latitude-velocity diagrams.}

\label{fig:pv}
\end{figure}
\clearpage

\begin{figure} \centering

\includegraphics[width=3in]{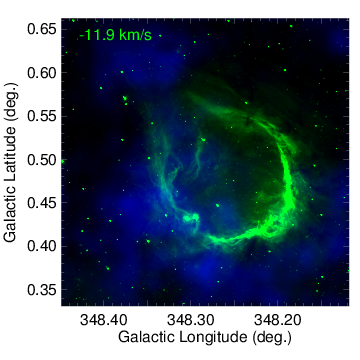}
\includegraphics[width=3in]{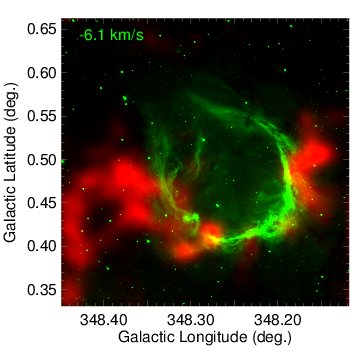}
\caption{The $-11.9$\,\kms\ (left) and $-6.1$\,\kms\ (right)
  \cor\ emission on top of 8.0\,\micron\ GLIMPSE data.  The
  8.0\,\micron\ emission shows the location of the PDR.}
\label{fig:twovel}
\end{figure}
\clearpage


\begin{figure} \centering

\includegraphics[width=6in]{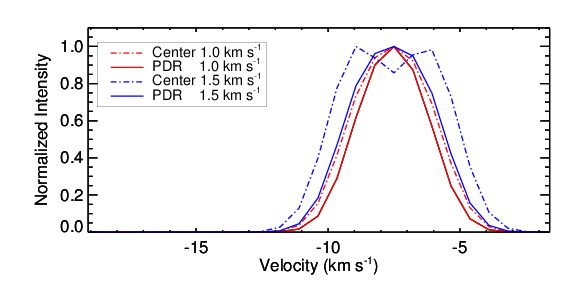}
\caption{Synthetic \cor\ spectra, generated using the radiative
  transfer code {\textsc MOLLIE}.  The curves show simulated spectra
  for a position along the PDR and toward the center of RCW\,120, for
  expansion velocities of 1.0 and 1.5\,\kms.  The 1.0\,\kms\ expansion
  speed is consistent with the observed spectra shown in
  Figure~\ref{fig:interior}, while the 1.5\,\kms\ expansion speed is
  not.}
\label{fig:model}
\end{figure}
\clearpage


\begin{figure} \centering

\includegraphics[scale=0.80]{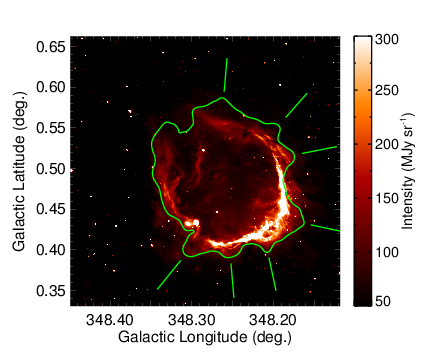}
\caption{Temperature enhancements outside the PDR of RCW\,120.
    The image is of GLIMPSE 8.0\,\micron\ data and the green outline
    shows the boundary of 22\,K {\it Herschel} dust temperatures from
    \citet{anderson12b} (hotter dust is inside this boundary).  We
    mark the locations of significant temperature enhancements outside
    the PDR with green lines.  These enhancements coincide with
    discontinuities in the PDR seen at 8.0\,\micron.}
\label{fig:enhancements}
\end{figure}
\clearpage


\begin{figure} \centering

\includegraphics[width=2.0in]{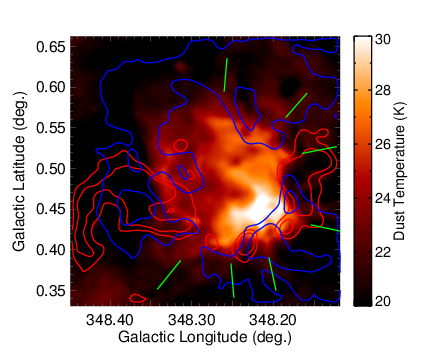}
\includegraphics[width=2.0in]{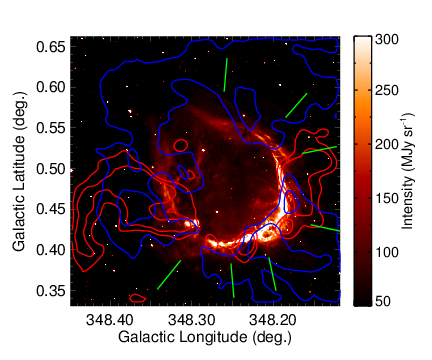}
\includegraphics[width=2.0in]{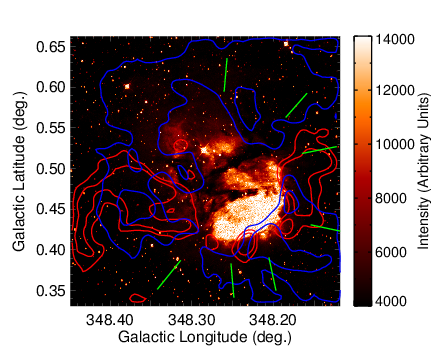}
\caption{(Left) Dust temperature map of RCW\,120 derived from
    {\it Herschel} data, showing temperature enhancements outside the
    PDR.  Red contours show $^{13}$CO emission at $-6.0$\,\kms
    (antenna temperature values of 1.4 and 2.0\,K), while blue
    contours show $^{13}$CO emission at $-8.9$\,\kms (antenna
    temperature values of 2 and 3\,K).  Locations of enhanced dust
    temperature are indicated with green lines, and correspond to
    deficits in the CO emission. (Middle) Same, for GLIMPSE
    8.0\,\micron\ data. (Right) Same, for SuperCosmos H$\alpha$
    data.}

\label{fig:holes}
\end{figure}
\clearpage

\end{document}